# Identification of microstructure from macroscopic measurement using inverse multiscale analysis

Anjan Mukherjee[1], Biswanath Banerjee[1]

[1] Department of Civil Engineering, Indian Institute of Technology Kharagpur, India
`anjan_iitkgp@iitkgp.ac.in`

**Abstract.** Most of the tailored materials are heterogeneous at the ingredient level. Analysis of those heterogeneous structures requires the knowledge of microstructure. With the knowledge of microstructure, multiscale analysis is carried out with homogenization at the micro level. Second-order homogenization is carried out whenever the ingredient size is comparable to the structure size. Therefore, knowledge of microstructure and its size is indispensable to analyzing those heterogeneous structures. Again, any structural response contains all the information of microstructure, like microstructure distribution, volume fraction, size of ingredients, etc. Here, inverse analysis is carried out to identify a heterogeneous microstructure from macroscopic measurement. Two-step inverse analysis is carried out in the identification process; in the first step, the macrostructure's length scale and effective properties are identified from the macroscopic measurement using gradient-based optimization. In the second step, those effective properties and length scales are used to determine the microstructure in inverse second-order homogenization.

## 1 Introduction

Classical continuum mechanics theory is well established and describes the deformation phenomena of versatile fundamental problems in different civil, mechanical, chemical, and metallurgical engineering fields. However, classical continuum mechanics theory fails to explain many size-dependent experimentally observed responses; for example, deformation around the nano-indentation point [19, 25, 12] and thickness-dependent deformation behavior of the nano-cantilever [15, 18].
In order to capture the size-dependent behavior, many nonlocal theories are considered, starting with [7]. These nonlocal theories [23, 24] consider additional strengthening due to strain gradients via length scales, and additional boundary conditions are also considered. Strain gradient elasticity by Aifantis [2, 27, 1] is one of the most popular nonlocal theories, considering a single length scale in the formulation. Application of strain gradient elasticity theory requires knowledge of the length scale, which represents the heterogeneity associated with the microstructure [4]. These length scales of identification require costly micro-level experiments [5, 15, 16, 17]. Another way of finding the length scales is to carry out the second-order homogenization [13] with the knowledge of microstructure images through costly micro-level scanning.

Heterogeneous structures can be analyzed using the multiscale method [9]. In the multiscale FE2 method [9], both the macro and microstructures are analyzed in finite elements with proper interaction. Homogenization is carried out at the microscale to obtain effective properties, and the homogenized properties are transferred to the macroscale. When the heterogeneity size is comparable to the size of the structure, second-order homogenization is carried out [13]. This second-order homogenization possesses effective properties, which contain the length scale associated with the microstructure.

The structural response depends on the heterogeneous microstructure. Therefore, it may be possible to identify the length scale, representing heterogeneity. In this work, the length scale of a heterogeneous microstructure is determined using macroscopic displacement measurements. Next, the length scale is used in inverse second-order homogenization to identify the possible microstructure.

## 2   Theory

### 2.1   Gradient elasticity theory and its implementation

Mindlin and Eshel [23, 24] proposed the gradient elasticity theory considering the contribution of strain gradient in strain energy density. Their formulation contained five material intrinsic length scales. Aifantis simplified the previous formulation with single length scales [2, 27, 1]. Here Aifantis' gradient elasticity theory is briefly reviewed. Virtual work principle related to the gradient elasticity can be written as

$$\int_V (\sigma : \delta\varepsilon + \tau \vdots \delta\eta) \ dV = \int_V [b \cdot \delta u] \ dV + \int_S [f \cdot \delta u + r \cdot D(\delta u)] \ dS \qquad (1)$$

$\sigma$ is the Cauchy stress, $\tau$ is the higher order stress, work conjugate to the strain gradient $\eta$, $\varepsilon$ is the strain tensor. The above weak form contains the boundary traction $f$ and higher order moment traction $r$, and $D(\cdot) = \boldsymbol{n} \cdot \boldsymbol{\nabla}(\cdot)$.

Aifantis proposed the following constitutive relation with single length scale ($l$)

$$\sigma_{ij} = \lambda \varepsilon_{kk} \delta_{ij} + 2\mu \varepsilon_{ij} \qquad (2a)$$

$$\tau_{kij} = l^2 \big(\lambda \eta_{kmm} \delta_{ij} + 2\mu \eta_{kij}\big) = l^2 \big(\lambda \varepsilon_{kk} \delta_{ij} + 2\mu \varepsilon_{ij}\big)_{,k} \qquad (2b)$$

The weak form of the equilibrium equation contains the second derivative of displacement. Therefore, $C^1$ type triangular elements, as proposed by [8], are used. The degrees of freedom at each node are displacements and their derivatives up to the second order. Hence, each node will have twelve degrees of freedom, and the total degree of freedom per element is thirty-six.

The weak form equation (1) can be written to the following discrete form with the help of nodal unknown displacement {**u**}

$$\int_V \{u\}^T \left([B_\varepsilon]^T[C][B_\varepsilon] + [B_\eta]^T[D][B_\eta]\right)\{\delta u\}\, dV = \int_S \{N\}^T\{t\}\{\delta u\}\, dS \quad (3)$$

## 2.2 Second-order homogenization

Second-order homogenization allows modeling a micro-structure with ingredient size comparable to macrostructural size. Length scale due to heterogeneity is automatically taken care of in second-order homogenization. Thus, second-order homogenization circumvents the requirement of considering any separate length scale in analysis.

In second-order homogenization, the RVE is solved to incorporate strain and strain gradient, which form the boundary of the RVE. The RVE is solved with a standard finite element considering a classical continuum solid. Upon solving the RVE, average stress and higher-order stress are transferred to the macrostructure. Second-order homogenization theory considering the effect of strain and strain gradient can be found in [14, 13, 11, 20].

In multi-scale modeling with higher-order homogenization, one RVE is solved at each Gauss point of the macrostructure. Being a gradient elastic medium, micro-structure sends the strain ($\varepsilon$) and strain gradient ($\eta$) information to the RVE. Upon solving, the RVE resultant stress ($\sigma$), double stress ($\tau$), and constitutive matrix **C**, **D** are transferred to the Gauss point. The multi-scale interaction is shown in Figure 1.

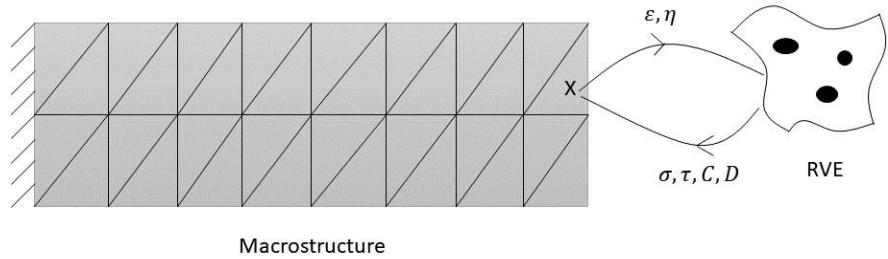

Figure 1: Micro-macro interaction

## 2.3 The inverse problem

In the inverse problem, unknown effective material properties and length scale of a porous cantilever beam are assessed in the first stage. In the second stage, the identified length scale and effective material properties are used to identify the microstructure.

**Length scale identification:**

The objective function to be minimized is the difference in measurements from the numerically simulated displacement. The objective functional to be minimized is

$$\psi_1 = \frac{1}{2} \frac{||[\Lambda]\{u(\alpha)\} - \{u^{exp}\}||^2}{||u^{exp}||^2}$$

(4)

where $[\Lambda]$ is the operator to bring down full field simulated displacement, $\{\hat{u}\}$ to discrete displacement corresponding to measured displacement, and $\{\mathbf{u}^{exp}\}$ is the experimental displacement. $\boldsymbol{\alpha}$ contains the information of effective Lame's constants and length scale, that is $\{\boldsymbol{\alpha}\} = \{\lambda^{eff}, \mu^{eff}, l^{eff}\}$. Gradient-based optimization is used to find the effective material properties and length scales.

**Microstructure identification:**

In the second step of the inverse problem, the volume fraction of the ingredients and the size of the inclusion is identified. The effective material properties and length scale found in the previous step are used to prepare the effective constitutive matrix $\mathbf{C}^{eff}$ and $\mathbf{D}^{eff}$. Now the inverse analysis is carried out to obtain the unknown volume fraction ($v_f$) and size of the inclusion ($\phi$) with known ingredient elastic properties. The objective function that is minimized in this step is the difference in norm between the calculated constitutive matrix and the effective constitutive matrix. The objective function can be written as

$$\psi_2 = \frac{1}{2} \frac{||C(\beta) - C^{eff}||^2}{||C^{eff}||^2} + \frac{1}{2} \frac{||D(\beta) - D^{eff}||^2}{||D^{eff}||^2},$$

(5)

where $\boldsymbol{\beta} = [\phi, v_f]$. The objective function is minimized with the help of gradient-based optimization. Gradient of the objective function is calculated using finite difference technique, that is

$$\frac{\partial \psi_2}{\partial \beta} = \frac{\psi_2(\beta + \delta\beta) - (\beta)}{\delta\beta}.$$

(6)

$\delta\boldsymbol{\beta}$ is the small change in the $\boldsymbol{\beta}$ value. The representative volume element (RVEs) are prepared in [22] considering a random sequential algorithm [21], and the finite element discretization is created with the help of Delaunay triangulation [10].

## 3. Result and discussion

Generally, a significant strain gradient effect is possessed by a cantilever beam with a point [3, 6]. Therefore, one cantilever is considered to identify its microstructure using the two-step inverse problem. The synthetic experimental displacement is prepared with some reference material properties and microstructure. In order to represent the measurement error associated with the experiment, the synthetic data is corrupted with random noise as

$$u_{noisy} = u_{exact}[1 + \gamma_n r] \qquad (7)$$

where, $\boldsymbol{u}_{noisy}$ is the corrupted displacements, $\boldsymbol{u}_{exact}$ is displacement calculated using FE2 formulation, $\gamma_n$ is the percentage of error and $r$ is the random number with a uniform distribution between $[-1,1]$. Top surface displacement is used as synthetic experimental data for the inverse problem. The general arrangement of the cantilever is shown in Figure 2. The cantilever is analyzed in FE2 [9] multiscale method with second-order homogenization [13] at the micro level. One porous microstructure of aluminum is used to prepare the synthetic displacement.

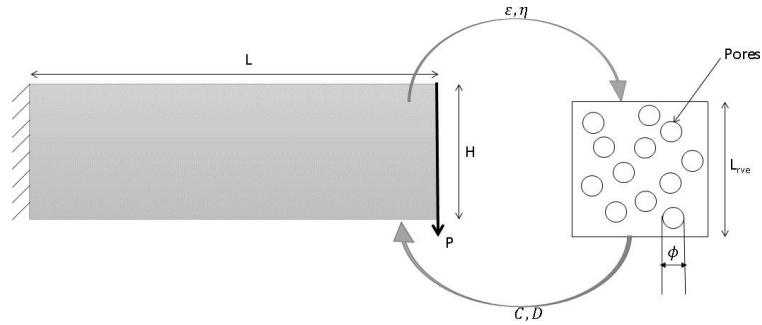

Figure 2: General arrangement of the cantilever

The beam has a length-to-depth ratio (L/H) equal to 3 with a depth (H) of 10 mm, and it is analyzed in plane strain gradient elasticity. The beam is discretized with a 75×25 number of grids, with each grid containing two triangular elements, and it is subjected to a point load of 1 kN at the tip. Considered material properties of aluminum are: Young's modulus of 70 GPa and Poisson's ratio of 0.3. The resulting Lame's parameters are λ = 40.38 GPa, μ = 26.92 GPa. The top displacement is corrupted with 5% noise according to equation (7).

The microstructure is prepared by distributing 15% of circular pores with a diameter of 0.3 mm in the matrix. The pores are randomly distributed in the matrix with the help of a random sequential algorithm [21]. The microstructure image is then discretized into finite elements with the help of Delaunay triangulation [10]. For a random microstructure, the size of RVE is generally considered to be 8-10 times the largest ingredient [26]. Here RVE size is assumed to be ten times the pore diameter, that is, 3 mm.

The first inverse problem is associated with identification of the unknown effective properties and length scale of the cantilever. The inverse problem was stated with the initial guess value of $\lambda_0 = 40.38$ GPa, $\mu_0 = 26.92$ GPa, and $l_0 = 3$ mm. The optimized length scales and material properties are $\lambda_{eff} = 24.22$ GPa, $\mu = 17.09$ GPa, and $l_{eff} = 0.93$ mm.

Next, these effective properties, along with known ingredient properties, are used in the second inverse problem to identify the microstructure. The guess value of this identification problem is $\phi = 0.1$ mm, and $v_f = 5$ %. Result of this inverse analysis is tabulated in Table 1. The evolution of RVE with iteration is presented in Figure 3. It is observed that the proposed inverse method is capable of capturing the microstructure when the ingredient properties are known in hand.

Table 1: Table showing the results of the micro-structure identification problem

| Guess Values | | Achieved values | | Reference values | |
|---|---|---|---|---|---|
| $\phi$ (mm) | $v_f$ (%) | $\phi$ (mm) | $v_f$ (%) | $\phi$ (mm) | $v_f$ (%) |
| 0.1 | 5 | 0.32 | 15.8 | 0.3 | 15 |

The optimization predicts the volume fraction and hole diameter within a few iterations. It is to note that an infinite number of distributions with the same inclusion size and volume fraction can possess similar displacement in the cantilever. Therefore, the identified microstructure is one of the representatives among them. It is indeed difficult to identify the distribution of the microstructure from only macro measurements.

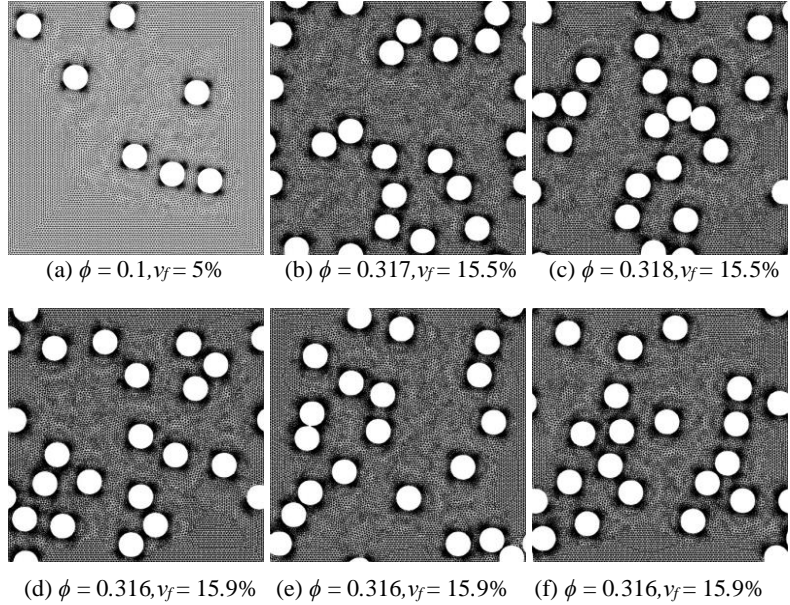

(a) $\phi = 0.1, v_f = 5\%$  (b) $\phi = 0.317, v_f = 15.5\%$  (c) $\phi = 0.318, v_f = 15.5\%$

(d) $\phi = 0.316, v_f = 15.9\%$  (e) $\phi = 0.316, v_f = 15.9\%$  (f) $\phi = 0.316, v_f = 15.9\%$

Figure 3: Evolution of RVE with iteration (a) Initial iteration (b) Iteration 1 (c) Iteration 2 and so on.

## 4. Conclusion

We have applied a gradient-based optimization technique to identify the length scale and effective material properties of a structure from macroscopic measurement. The identified length scale is used in inverse homogenization to identify the microstructure. The inverse homogenization is carried out with known ingredient properties and assumed RVE size to inclusion size ratio. The effectiveness of the proposed identification technique is established by identifying a porous microstructure of a cantilever. However, the identification technique can also identify microstructure with inclusion instead of pores. Multiple microstructures with equal inclusion size and volume fraction can possess equal homogenized properties. Therefore, the identified microstructure is representative of one of them. The inverse problem considers simple circular pores. In reality, the actual distribution and shape of pores may differ from the identified one as it is almost impossible to identify those from only macroscopic measurements.